\documentclass[cup7b]{cupbook}
\def\be{\begin{equation}}
\def\ee{\end{equation}}
\def\beq{\begin{eqnarray}}
\def\eeq{\end{eqnarray}}

\def\vp{\varphi}

\def\N{{\cal N}}

\def\({\left (}
\def\){\right )}
\def\[{\left [}
\def\[{\right ]}
\def\S{{\cal S}}

\begin{document}

\newtheorem{theorem}{Theorem}[chapter]
\newcommand{\blackboard}{\bf }

\pagenumbering{arabic}

\author[Gary T. Horowitz and Joseph Polchinski]{GARY T. HOROWITZ\\
University of California at Santa Barbara \and JOSEPH POLCHINSKI\\
KITP, University of California at Santa Barbara}
\chapter{Gauge/gravity duality} 

\begin{abstract}
We review the emergence of gravity from gauge theory in the context of AdS/CFT duality.  We discuss the evidence for the duality, its lessons for gravitational physics, generalizations, and open questions.
\end{abstract}

\section{Introduction}

{\it Assertion: Hidden within every non-Abelian gauge theory, even within the weak and strong nuclear interactions, is a theory of quantum gravity.}

This is one implication of AdS/CFT duality.  It was discovered by a circuitous route, involving in particular the relation between black branes and D-branes in string theory.  It is an interesting exercise, however, to first try to find a path from gauge theory to gravity as directly as possible.  Thus let us imagine that we know a bit about gauge theory and a bit about gravity but nothing about string theory, and ask, how are we to make sense of the assertion? 

One possibility that comes to mind is that the spin-two graviton might arise as a composite of two spin-one gauge bosons.  This interesting idea would seem to be rigorously excluded by a no-go theorem of Weinberg \& Witten (1980).  The Weinberg-Witten theorem appears to assume nothing more than the existence of a Lorentz-covariant energy momentum tensor, which indeed holds in gauge theory.  The theorem does forbid a wide range of possibilities, but (as with several other beautiful and powerful no-go theorems) it has at least one hidden assumption that seems so trivial as to escape notice, but which later developments show to be unnecessary.  The crucial assumption here is that the graviton moves in the same spacetime as the gauge bosons of which it is made!

The clue to relax this assumption comes from the holographic principle ('t Hooft, 1993, and Susskind, 1995), which suggests that a gravitational theory should be related to a non-gravitational theory in {\it one fewer} dimension.  In other words, we must find within the gauge theory not just the graviton, but a fifth dimension as well: the physics must be local with respect to some additional hidden parameter.  Several hints suggest that the role of this fifth dimension is played by the {\it energy scale} of the gauge theory.  For example, the renormalization group equation is local with respect to energy: it is a nonlinear evolution equation for the coupling constants as measured at a given energy scale.\footnote{This locality was emphasized to us by Shenker, who credits it to Wilson.}

In order to make this precise, it is useful to go to certain limits in which the five-dimensional picture becomes manifest; we will later return to the more general case.  Thus we consider four-dimensional gauge theories with the following additional properties:
\begin{itemize}
\item
{\it Large $N_c$.}
While the holographic principle implies a certain equivalence between four- and five-dimensional theories, it is also true that in many senses a higher dimensional theory has more degrees of freedom; for example, the one-particle states are labeled by an additional momentum parameter.  Thus, in order to find a fifth dimension of macroscopic size, we need to consider gauge theories with many degrees of freedom.  A natural limit of this kind was identified by 't Hooft (1974): if we consider $SU(N_c)$ gauge theories, then there is a smooth limit in which $N_c$ is taken large with the combination $g^2_{\rm YM} N_c$ held fixed.
\item
{\it Strong coupling.}  Classical Yang-Mills theory is certainly not the same
as classical general relativity.  If gravity is to emerge from gauge theory,
we should expect that it will be in the limit where the gauge fields
are strongly quantum mechanical, and the gravitational degrees of
freedom arise as effective classical fields. Thus we must consider the
theory with large Õt Hooft parameter $g^2_{\rm YM} N_c$.
\item
{\it Supersymmetry.} This is a more technical assumption, but it is a natural
corollary to the previous one. Quantum field theories at strong
coupling are prone to severe instabilities; for example, particle-antiparticle pairs can appear spontaneously, and their negative potential energy would exceed their positive rest and kinetic energies.  Thus, QED with fine structure constant much greater than 1 does not exist, even as an effective
theory, because it immediately runs into an instability in the ultraviolet (known as the Landau pole).  The Thirring model provides a simple solvable illustration
of the problem: it exists only below a certain critical coupling (Coleman, 1975).  Supersymmetric theories
however have a natural stability property, because the Hamiltonian is
the square of a Hermitean supercharge and so bounded below. Thus it
is not surprising that most examples of field theories with interesting
strong coupling behavior (i.e. dualities) are supersymmetric.  We will therefore start by assuming supersymmetry, but after understanding this case we can work back to the nonsupersymmetric case.
\end{itemize}

We begin with the most supersymmetric
possibility, ${\cal N} = 4$ $SU(N_c)$ gauge theory, meaning that there
are four copies of the minimal $D = 4$ supersymmetry algebra. 
The
assumption of ${\cal N} = 4$ supersymmetry has a useful bonus in that the
beta function vanishes, the coupling does not run. Most gauge theories
have running couplings, so that the strong coupling required by
the previous argument persists only in a very narrow range of energies,
becoming weak on one side and blowing up on the other. In the
${\cal N} = 4$ gauge theory the coupling remains strong and constant over
an arbitrarily large range, and so we can have a large fifth
dimension.

The vanishing beta function implies that the classical conformal invariance
of the Yang-Mills theory survives quantization: it is a conformal field theory (CFT).  In particular,
the theory is invariant under rigid scale transformations $x^\mu \to \lambda x^\mu$ for $\mu  = 0, 1, 2, 3$. Since we are associating the fifth coordinate $r$ with energy
scale, it must tranform inversely to the length scale, $r \to r/\lambda$. The most
general metric invariant under this scale invariance and the ordinary
Poincar\'e symmetries is
\begin{equation}
ds^2 = \frac{r^2}{\ell'^2} \eta_{\mu\nu} dx^\mu dx^\nu   + \frac{\ell^2}{r^2} dr^2 \label{adsmet}
\end{equation}
for some constants $\ell$ and $\ell'$; by a multiplicative redefinition of $r$ we
can set $\ell' = \ell$. Thus our attempt to make sense of the assertion at
the beginning has led us (with liberal use of hindsight) to the following
conjecture: $D = 4$, ${\cal N} = 4$, $SU(N_c)$ gauge theory is equivalent to a
gravitational theory in five-dimensional anti-de Sitter (AdS) space. Indeed,
this appears to be true. In the next section we will make this statement
more precise, and discuss the evidence for it.  In the final section we will discuss various lessons for quantum gravity, generalizations, and open questions.

\section{AdS/CFT duality}

Let us define more fully the two sides of the duality.\footnote
{This subject has a vast literature, and so we will be able to cite only a few particularly pertinent references.  We refer the reader to the review Aharony \mbox{\it et al.} (2000) for a more complete treatment.}
  The gauge theory can be written in a compact way by starting with the $D=10$ Lagrangian density for an $SU(N_c)$ gauge field and a 16 component Majorana-Weyl spinor, both in the adjoint ($N_c \times N_c$ matrix) representation:
\begin{equation}
{\cal L} = \frac{1}{2 g_{\rm YM}^2} {\rm Tr} (F_{\mu\nu} F^{\mu\nu}) + i {\rm Tr} (\bar\psi \gamma^\mu D_\mu \psi)\ .
\end{equation}
This Lagrangian preserves 16 supersymmetries, the smallest algebra in $D=10$.  Now dimensionally reduce to $D=4$, meaning that we define all fields to be independent of the coordinates $x^{4}, \ldots, x^{9}$.  The ten-dimensional gauge field separates into a four-dimensional gauge field and six scalars $\varphi^i$, and the ten-dimensional spinor separates into four four-dimensional Weyl spinors.

On the other side of the duality, we must consider not just gravity but its supersymmetric extension, to match what we have in the gauge theory.  The necessary theory is IIB supergravity.  This theory too is most naturally formulated in $D=10$, where its fields includes the metric, two scalars $\Phi$ and $C$, two two-form potentials $B_{MN}$ and $C_{MN}$, a four-form potential $C_{MNPQ}$ whose five-form field strength is self-dual, and fermionic partners (including the gravitino) as required by supersymmetry.  This ten-dimensional theory has a solution with spacetime geometry $AdS_5 \times S^5$.  In fact, one finds that it is this full ten-dimensional theory that arises in the strong-coupling limit of the gauge theory.  There emerges not only the fifth dimension required by holography, but five more.  The additional five dimensions, which can be thought of as arising from the  scalars $\varphi^i$, form a compact five-sphere.

On both sides of the duality we have started in $D=10$, because this is the natural dimensionality for this supersymmetry algebra.  On the gauge side, however, this was just a device to give a compact description of the Lagrangian; the field theory lives in four dimensions.  On the gravity side, the quantum theory is fully ten-dimensional, not just a dimensional reduction.  These statements follow from comparison of the space of states, or from the original Maldacena argument, as we will shortly explain.

The claim that a four-dimensional gauge theory gives rise to a ten-dimensional gravitational theory is remarkable.  One sign that it is not completely crazy comes from
comparing the symmetries. The  $D = 4$, ${\cal N} = 4$, $SU(N_c)$ super-Yang-Mills theory has an $SO(4,2)$ symmetry coming from conformal invariance and an $SO(6)$ symmetry coming
from rotation of the scalars. This agrees with the geometric
symmetries of $AdS_5\times S^5$.  On both sides there are also 32 supersymmetries.  Again on the gravitational side these are geometric, arising as Killing spinors on the $AdS_5 \times S^5$ spacetime.  On the gauge theory side they include the 16 `ordinary' supersymmetries of the $\N = 4$ algebra, and 16 additional supersymmetries as required by the conformal algebra.

The precise (though still not fully complete) statement is that the IIB supergravity theory in a space whose geometry is asymptotically $AdS_5\times S^5$ is dual to the $D = 4$, ${\cal N} = 4$, $SU(N_c)$ gauge theory.  The metric~(\ref{adsmet}) describes only a Poincar\'e patch of AdS spacetime, and the gauge theory lives on ${\bf R}^4$.  It is generally more natural to consider the fully extended global AdS space, in which case the dual gauge theory lives on $S^3\times {\bf R} $.  In each case the gauge theory lives on the conformal boundary of the gravitational spacetime ($r \to\infty$ in the Poincar\'e coordinates), which will give us a natural dictionary for the observables.

The initial checks of this duality concerned perturbations of $AdS_5\times S^5$.
It was shown that all linearized supergravity states have corresponding states in the gauge theory (Witten, 1998a).  In particular, the global time translation in the bulk is identified with time translation in the field theory, and the energies of states in the field theory and string theory agree. For perturbations of $AdS_5\times S^5$, one can reconstruct the background
spacetime from the gauge theory as follows.  Fields on $S^5$ can be decomposed into spherical
harmonics, which can be described as symmetric traceless tensors on ${\bf R}^6:$
$T_{i\cdots j} X^i \cdots X^j$.  Restricted to the unit sphere one gets a basis of functions.  Recall that the gauge theory has six scalars and the SO(6) symmetry
of rotating the $\vp^i$. So the operators $T_{i\cdots j} \vp^i\cdots \vp^ j$
give information about position on $S^5$.  Four of the remaining directions are explicitly present in the gauge theory, and 
 the radial direction corresponds to the energy scale in the gauge theory.

In the gauge theory the expectation values of local operators (gauge invariant products of the ${\cal N} = 4$ fields and their covariant derivatives) provide one natural set of observables.  It is convenient to work with the generating functional for these expectation values by shifting the Lagrangian density 
\begin{equation}
\textstyle {\cal L}(x) \to {\cal L}(x) + \sum_I J_I(x) {\cal O}_I(x)\ , \label{deform}
\end{equation}
where ${\cal O}_I$
is some basis of local operators and $ J_I(x)$ are arbitrary functions.  Since we are taking products of operators at a point, we are perturbing the theory in the ultraviolet, which according to the energy-radius relation maps to the AdS boundary.  Thus the duality dictionary relates the gauge theory generating functional to a gravitational theory in which the boundary conditions at infinity are perturbed in a specified way~(Gubser \mbox{\it et al.}, 1998, and Witten, 1998a).  As a further check on the duality, all three-point interactions were shown to agree (Lee \mbox{\it et al.}, 1998).

The linearized supergravity excitations map to gauge invariant states of the gauge bosons, scalars, and fermions, but in fact only to a small subset of these; in particular, all the supergravity states live in special small multiplets of the superconformal symmetry algebra.  Thus the dual to the gauge theory contains much more than supergravity.  The identity of the additional degrees of freedom becomes particularly clear if one looks at highly boosted states, those having large angular momentum on $S^5$ and/or $AdS_5$ (Berenstein \mbox{\it et al.}, 2002, and Gubser  \mbox{\it et al.}, 2002).  The fields of the gauge theory then organize naturally into one-dimensional structures, coming from the Yang-Mills large-$N_c$ trace: they correspond to the excited states of strings.   In some cases, one can even construct a two dimensional sigma model directly from the gauge
theory and show that it agrees (at large boost) with the sigma model describing strings moving in $AdS_5\times S^5$ (Kruczenski, 2004).

Thus, by trying to make sense of the assertion at the beginning, we are forced to `discover'  string theory.  We can now state the duality in its full form (Maldacena, 1998a): 
\begin{center}
{\it Four-dimensional ${\cal N}=4$ supersymmetric
$SU(N_c)$ gauge theory is equivalent to IIB string theory with $AdS_5 \times S^5$ boundary conditions.}
\end{center}
\noindent
The need for strings (though not the presence of gravity\mbox{!}) was already anticipated by 't Hooft (1974), based on the planar structure of the large-$N_c$ Yang-Mills perturbation theory; the AdS/CFT duality puts this into a precise form.  
It also fits with the existence of another important set of gauge theory observables, the one-dimensional Wilson loops.  The Wilson loop can be thought of as creating a string at the $AdS_5$ boundary, whose world-sheet then extends into the interior (Maldacena, 1998b, and Rey \& Yee, 2001).  See also Polyakov (1987, 1999) for other perspectives on gauge/string duality and the role of the fifth dimension.

We now drop the pretense of not knowing string theory, and outline the original argument for the duality in Maldacena (1998a).  He considered a stack of $N_c$ parallel D3-branes on top of each other.  Each D3-brane couples to gravity with a strength proportional to the dimensionless string coupling $g_s$, so the distortion of the metric by the branes is proportional to $g_s N_c$. When $g_s N_c \ll 1$ the spacetime is nearly flat and there are two types of string excitations. There are open strings on the brane whose low energy modes are described by a $U(N_c)$ gauge theory.  There are also closed strings away from the brane. When $g_s N_c\gg 1$, the
backreaction is important and the metric describes an extremal black 3-brane.
This is a generalization of a black hole appropriate for a three dimensional extended object. It is extremal with respect to the charge carried by the 3-branes, which sources the five form $F_5$.
Near the horizon, the spacetime becomes a product of $S^5$ and $AdS_5$.  (This is directly
analogous to the fact that near the horizon of an extremal Reissner-Nordstrom black hole, the spacetime is $AdS_2\times S^2$.)  String states near the horizon are strongly redshifted and have very low energy as seen asymptotically.  In a certain low energy
limit, one can decouple these strings from the strings in the asymptotically flat region.  At
weak coupling, $g_s N_c\ll 1$, this same limit decouples the excitations of the 3-branes from the closed strings.  Thus the low energy decoupled physics is described by the gauge theory at small $g_s$ and by the $AdS_5 \times S^5$ closed string theory at large $g_s$, and the simplest conjecture is that these are the same theory as seen at different values of the coupling.\footnote{The $U(1)$ factor in $U(N_c) = SU(N_c) \times U(1)$ also decouples: it is Abelian and does not feel the strong gauge interactions.}
This conjecture resolved a puzzle, the fact that very different gauge theory and gravity calculations were found to give the same answers for a variety of string-brane interactions.

In the context of string theory we can relate the parameters on the two sides of the duality.  In the gauge theory we have $g_{\rm YM}^2$ and $N_c$.  The known D3-brane Lagrangian determines the relation of couplings, $g_{\rm YM}^2 = 4\pi g_s$.  Further, each D3-brane is a source for the five-form field strength, so on the string side $N_c$ is determined by $\int_{S^5} F_5$; this integrated flux is quantized by a generalization of Dirac's argument for quantization of the flux $\int_{S^2} F_2$ of a magnetic monopole.  The supergravity field equations give a relation between this flux and the radii of curvature of the $AdS_5$
and $S^5$ spaces, both being given by
\be\label{adsradius}
\ell = (4\pi g_s N_c)^{1/4} \ell_s\ .
\ee
Here $\ell_s$ is the fundamental length scale of string theory, related to the string tension $\mu$ by $\mu^{-1} = 2\pi \ell_s^{2}$.  Notice that the spacetime radii are large in string units (and so the curvature is small) precisely when the 't Hooft coupling $4\pi g_s N_c = g_{\rm YM}^2 N_c$ is large, in keeping with the heuristic arguments that we made in the introduction.
It is also instructive to express the AdS radius entirely in gravitational variables.  The ten-dimensional gravitational coupling is $G \sim g_s^2 \ell_s^8$, up to a numerical constant.  Thus
\begin{equation}
\ell \sim N_c^{1/4} G^{1/8}\ , \quad G \sim \frac{\ell^8}{N_c^2}\ . \label{adsplanck}
\end{equation}
In other words, the AdS radius is $N_c^{1/4}$ in Planck units, and the gravitational coupling is $N_c^{-2}$ in AdS units.

\section{Lessons, generalizations, and open questions}

\subsection{Black holes and thermal physics}

The fact that black holes have thermodynamic properties is one of the most striking features of classical and quantum gravity.  In the context of AdS/CFT duality, this has a simple realization: in the dual gauge theory the black hole is just a hot gas of gauge bosons, scalars, and fermions, the gauge theory degrees of freedom in equilibrium at the Hawking temperature.

A  black hole in $AdS_5$ is described by the Schwarzschild AdS geometry
\be\label{adsbh}
ds^2 = -\({r^2\over \ell^2} +1 - {r_0^2\over r^2}\) dt^2 + \({r^2\over \ell^2}
+1 - {r_0^2\over r^2} \)^{-1}
dr^2 + r^2 d\Omega_3 \ .
\ee
Denoting the Schwarzschild radius by $r_+$, 
the Hawking temperature of this black hole is
$
T_H = {(\ell^2 + 2r_+^2)/ 2\pi r_+\ell^2}.
$
When $r_+ \gg\ell$, the Hawking
temperature is large, $T_H \sim r_+/\ell^2$. 
This is quite different from a large
black hole in asymptotically flat spacetime which has $T_H \sim 1/r_+$.
The gauge theory
description is just a thermal state at the same temperature $T_H$.

Let us compare the entropies in the two descriptions.  It is difficult to calculate the field
theory entropy at strong coupling, but at weak coupling, we have of
order $N_c^2$ degrees of freedom, on a three sphere of radius $\ell$ at 
temperature $T_H$ and hence 
\be\label{entropy}
S_{\rm YM} \sim N_c^2 T_H^3 \ell^3.
\ee
On the string theory side, the solution is the product of (\ref{adsbh})
and an $S^5$ of radius $\ell$. 
So recalling that $G\sim g_s^2\ell_s^8$ in ten dimensions and dropping factors
of order unity, the Hawking-Bekenstein entropy of this black hole is
\be 
S_{BH}={A\over 4G}\sim {r_+^3 \ell^5\over g_s^2 \ell_s^8}\sim {T_H^3\ell^{11}
\over g^2_s\ell_s^8}
\sim N_c^2 T_H^3 \ell^3
\ee
where we have used (\ref{adsradius}) in the last step. The agreement
with (\ref{entropy}) shows that the field theory has enough states to
reproduce the entropy of large black holes in $AdS_5$.  

On the gauge theory side, the scaling of the entropy as $T_H^3$ is just dimensional analysis for a massless field theory in 3+1 dimensions.  That the ten-dimensional string theory produces the same behavior is a surprising consequence of the AdS geometry.  The factor of $N_c^2$ similarly just counts explicit degrees of freedom on the gauge theory side, while on the string side it comes from the scaling of the horizon area.

Putting in all the numerical factors one finds that $S_{BH} = {3\over 4} S_{\rm YM}$ 
(Gubser \mbox{\it et al.}, 1996).  The numerical disagreement is not surprising, as the Yang-Mills calculation is for an ideal gas, and at large $g_s$ the Yang-Mills degrees of freedom are interacting.  Thus one expects a relation of the form $S_{BH} = f(g_s N_c) S_{\rm YM, ideal}$, where $f(0) = 1$; the above calculation implies that $f(\infty) = \frac{3}{4}$.  We do not yet have a quantitative understanding of the value $\frac34$, but the first correction has been calculated both at weak and strong coupling and is consistent with $f(g_s N_c)$ interpolating in a rather smooth way.  

Hawking \& Page (1983) showed that for thermal AdS boundary conditions  there is a phase transition:  below a transition temperature of order $1/\ell$ the dominant configuration is not the black hole but a gas of particles in  AdS space.  The low temperature geometry has no horizon and so its entropy comes only from the ordinary statistical mechanics of the gas.  The same transition occurs in the gauge theory (Witten, 1998b).  The ${\cal N}=4$ gauge theory on $S^3$ has an analog of a confinement transition.  At low temperature one has a thermal ensemble of gauge-invariant degrees of freedom, whose entropy therefore scales as $N_c^0$, and at high temperature one has the $N_c^2$ behavior found above --- the same scalings as on the gravitational side.

There is another test one can perform with the gauge theory at finite temperature. At long wavelengths, one can use a hydrodynamic approximation and think of this as a fluid (for a recent overview see Kovtun \mbox{\it et al.}, 2003). It is
then natural to ask:  What is the speed of sound waves? Conformal invariance implies that
the stress energy tensor is traceless, so $p=\rho/3$ which implies
that $v = 1/\sqrt 3$. The question is: Can you derive this sound speed
from the AdS side? This would seem to be difficult since the bulk does not
seem to have any preferred speed other than the speed of light. But recent
work has shown that the answer is yes. 

The AdS/CFT duality also gives an interesting perspective on the black hole membrane paradigm (Thorne \mbox{\it et al.}, 1986).  The black hole horizon is known to have many of the properties of a dissipative system.  On the dual side it {\it is} a dissipative system, the hot gauge theory.
One can thus compute such hydrodynamic quantities
such as the shear viscosity.  These are hard to check since they are difficult to calculate directly in the strongly coupled thermal gauge theory, but, rather remarkably, the numerical agreement with the observed properties of the real quark-gluon plasma at RHIC is better than for conventional field theory calculations (for a discussion see Blau, 2005).

There is also a field theory interpretation of black hole quasinormal modes (Horowitz \& Hubeny 2000). A perturbation of the black hole decays with a characteristic time set by the imaginary part of the lowest quasinormal 
mode. This should correspond to the timescale for the gauge theory to return to thermal equilibrium. One can show that the quasinormal mode frequencies are poles in the retarded Green's function of a certain operator in the gauge theory. The particular operator depends on the type of field used to perturb the black hole (Kovtun \& Starinets, 2005).

Finally, consider the formation and evaporation of a small black hole in a spacetime which is asymptotically $AdS_5\times S^5$.  By the AdS/CFT correspondence, this process is described by ordinary unitary evolution in the gauge theory. So black hole evaporation does not violate quantum mechanics: information is preserved.  This also provides an indirect argument against the existence of a `bounce' at the black hole singularity, because the resulting disconnected universe would presumably carry away information.

\subsection{Background independence and emergence}

The AdS/CFT system is entirely embedded in the framework of quantum mechanics.  On the gauge theory side we have an explicit Hamiltonian, and states which we can think of as gauge invariant functionals of the fields.  Thus the gravitational theory on the other side is quantum mechanical as well.  In particular the metric fluctuates freely except at the AdS boundary.  One is not restricted to perturbations about a particular background.

This is clearly illustrated by a rich set of examples  which provide a detailed map  between a
class of nontrivial asymptotically $AdS_5\times S^5$ supergravity solutions 
and a class of states in the gauge theory (Lin \mbox{\it et al.}, 2004).
These states and geometries both preserve half of the supersymmetry of $AdS_5\times S^5$ itself. On the field theory side, one restricts to fields that are independent of $S^3$ and hence
reduce to $N_c \times N_c$ matrices. In fact, all the states are created by a
single complex matrix, so can be described by a one-matrix model.  This
theory can be quantized exactly in terms of free fermions, and the states can be labeled by a arbitrary closed curve (the Fermi surface) on a plane. On the gravity side, one
considers solutions to ten dimensional supergravity involving just the metric
and self-dual five form $F_5$. The field equations are  simply $dF_5=0$ and
\be\label{fieldeq}
R_{MN} = F_{MPQRS}{F_N}^{PQRS}
\ee
There exists a large class
of stationary solutions to (\ref{fieldeq}), which have an
$SO(4)\times SO(4)$ symmetry and can be obtained by solving a linear equation.
These solutions are nonsingular, have no event horizons, but can have 
complicated topology. They are also labeled by arbitrary closed curves
on a plane. This provides a precise way to map states in the field
theory into bulk geometries.  Only for some ``semi-classical" states
is the curvature below the Planck scale everywhere, but the matrix/free fermion description readily describes all the states, of all topologies, within a single Hilbert space.

Thus the gauge theory gives a representation of quantum gravity that is background independent almost everywhere ---- that is, everywhere except the boundary.  Conventional string perturbation theory constructs string amplitudes as an asymptotic expansion around a given spacetime geometry; here we have an exact quantum mechanical construction for which the conventional expansion generates the asymptotics.  All local phenomena of quantum gravity, such as formation and evaporation of black holes, the interaction of quanta with Planckian energies, and even transitions that change topology, are described by the gauge theory.  However, the boundary conditions do have the important limitation that most cosmological situations, and most compactifications of string theory, cannot be described; we will return to these points later.    

To summarize, AdS/CFT duality is an example of emergent gravity, emergent spacetime, and emergent general coordinate invariance.  But it is also an example of emergent strings!  We should note that the terms `gauge/gravity duality' and `gauge/string duality' are often used, both to reflect these emergent properties and also the fact that (as we are about the see) the duality generalizes to gravitational theories with certain other boundary conditions, and to field theories that are not conformally invariant.

Let us expand somewhat on the emergence of general coordinate invariance.  The AdS/CFT duality is a close analog to the phenomenon of emergent gauge symmetry (e.g. D'Adda  \mbox{\it et al.}, 1978, and Baskaran \& Anderson, 1988).  For example, in some condensed matter systems in which the starting point has only electrons with short-ranged interactions, there are phases where the electron separates into a new fermion and boson,
\begin{equation}
e(x) = b(x) f^\dagger(x)\ .
\end{equation}
However, the new fields are redundant: there is a gauge transformation $b(x) \to e^{i\lambda(x)} b(x)$,
$f(x) \to  e^{i\lambda(x)} f(x)$, which leaves the physical electron field invariant.  This new gauge invariance is clearly emergent: it is completely invisible in terms of the electron field appearing in the original description of the theory.\footnote{This `statistical' gauge invariance is not to be confused with the ordinary electromagnetic gauge invariance, which does act on the electron.}
Similarly, the gauge theory variables of AdS/CFT are trivially invariant under the bulk diffeomorphisms, which are entirely invisible in the gauge theory (the gauge theory fields do transform under the asymptotic symmetries of $AdS_5 \times S^5$, but these are ADM symmetries, not gauge redundancies).  Of course we can always in general relativity introduce a set of gauge-invariant observables by setting up effectively a system of rods and clocks, so to this extent the notion of emergence is imprecise, but it carries the connotation that the dynamics can be expressed in a simple way in terms of the invariant variables, as  the case in AdS/CFT.\footnote{Note that on the gauge theory side there is still the ordinary Yang-Mills gauge redundancy, which is more tractable than general coordinate invariance (it does not act on spacetime).  In fact in most examples of duality there are gauge symmetries on both sides and these are unrelated to each other: the duality pertains only to the physical quantities.}
 
\subsection{Generalizations}

Thus far we have considered only the most well-studied example of gauge/gravity duality:
$D=4$, ${\cal N}=4$, Yang-Mills $\Leftrightarrow$ string theory with $AdS_5 \times S^5$ boundary conditions.  Let us now ask how much more general this phenomenon is (again, for details see the review by Aharony \mbox{\it et al.}, 2000).

First, we imagine perturbing the theory we have already studied, adding additional terms (such as masses for some of the fields) to the gauge theory action.  This is just a special case of the modification~(\ref{deform}), such that the functions $J_I(x) = g_I$ are independent of position.  Thus we already have the dictionary, that the the dual theory is given by IIB string theory in a spacetime with some perturbation of the $AdS_5 \times S^5$ boundary conditions.  
 
In general, the perturbation of the gauge theory will break conformal invariance, so that the physics depends on energy scale.  In quantum field theory
there is a standard procedure for integrating out high energy
degrees of freedom and obtaining an effective theory at low energy.
This is known as renormalization group (RG) flow. If one starts with a 
conformal field theory at high energy, the RG flow is trivial. The low
energy theory looks the same as the high energy theory. This is because
there is no intrinsic scale.  But if we perturb the theory, the RG flow is nontrivial and we obtain a different theory at low energies.  There are two broad possibilities: either some degrees of freedom remain massless and we approach a new conformal theory at low energy, or all fields become massive and the low energy limit is trivial.

Since the energy scale corresponds to the radius, this RG flow in the boundary field theory should correspond to radial dependence in the bulk.  Let us expand a bit on the relation between radial coordinate and energy (we will make this argument in Poincar\'e coordinates, since the perturbed gauge theories are usually studied on ${\bf R}^4$).  The AdS geometry~(\ref{adsmet}) is {\it warped}: in Poincare coordinates, the four flat dimensions experience a gravitational redshift that depends on fifth coordinate, just as in Randall-Sundrum compactification.  Consequently the conserved Killing momentum $p_\mu$ (Noether momentum in the gauge theory) is related to the local inertial momentum $\tilde p_\mu$ by
\begin{equation}
p_\mu = \frac{r}{\ell} \tilde p_\mu\ .  \label{warp}
\end{equation}
A state whose local inertial momenta are set by the characteristic scale $1/\ell$ therefore has a Killing momentum $p_\mu \sim r/\ell^2$, displaying explicitly the mapping between energy/momentum scale and radius.

Given a perturbation that changes the boundary conditions, AdS is no longer a solution and we must 
solve Einstein's equation to find the correct solution.  Just as in the gauge theory there are two possibilities: either we approach a new AdS solution at small radius (with, in general, a different
radius of curvature), or the small radius geometry is cut off in such a way that the warp factor (which is $r/\ell$ in AdS spacetime) has a lower bound.  The former clearly corresponds to a new conformal theory, while the latter would imply a mass gap, by the argument following eq.~(\ref{warp}).  In the various examples, one finds that the nature of the solution correctly reflects the low energy physics as expected from gauge theory arguments; there is also more detailed numerical agreement (Freedman  \mbox{\it et al.}, 1999).  So the classical Einstein
equation knows a lot about RG flows in quantum field theory.  

A notable example is the case where one gives mass to all the scalars and fermions, leaving only the gauge fields massless in the Lagrangian.  One then expects the gauge theory to flow to strong coupling and produce a mass gap, and this is what is found in the supergravity solution.  Further, the gauge theory should confine, and indeed in the deformed geometry a confining area law is found for the Wilson loop (but still a perimeter law for the 't Hooft loop, again as expected).  In other examples one also finds chiral symmetry breaking, as expected in strongly coupled gauge theories (Klebanov \& Strassler, 2000).

As a second generalization, rather than a deformation of the geometry we can make a big change, replacing $S^5$ with any other Einstein space; the simplest examples would be $S^5$ identified by some discrete subgroup of its $SO(6)$ symmetry.  The product of the Einstein space with $AdS_5$ still solves the field equations (at least classically), so there should
be a conformally invariant dual.  These duals are known in a very large class of examples; characteristically they are quiver gauge theories, a product of $SU(N_1) \times \ldots \times SU(N_k)$ with matter fields transforming as adjoints and bifundamentals (one can also get orthogonal and symplectic factors).

As a third generalization, we can start with D$p$-branes for other values of $p$, or combinations of branes of different dimensions.  These lead to other examples of gauge-gravity duality for field theories in various dimensions, many of which are nonconformal.  The case $p=0$ is the BFSS matrix model, although the focus in that case is on a different set of observables, the scattering amplitudes for the D0-branes themselves.
A particularly interesting system is D1-branes plus D5-branes, leading to the near-horizon geometry $AdS_3\times S^3\times T^4$.
This case has at least one advantage over $AdS_5\times S^5$.  The entropy of large
black holes can now be reproduced exactly, including the numerical coefficient. This is related to the fact that a black hole in $AdS_3$ is a BTZ black hole
which is locally $ AdS_3$ everywhere. Thus when one extrapolates to small
coupling, one does not modify the geometry with higher curvature corrections.

We have discussed modifications of the gauge theory's Hamiltonian, its spectrum, and even its dimensionality.  Many of these break the theory's conformal symmetry and some or all of its supersymmetry (with all of it broken the stability is delicate, but possible).  Thus we can relax the assumption of supersymmetry, as promised earlier.  If we start with a nonsupersymmetric gauge theory, do we get a gravitational theory without supergravity (and maybe without strings)?  Apparently not.  When we change the dynamics of the gauge theory, we do {\it not} change the local dynamics of the gravitational theory, i.e.~its equation of motion, but only its boundary conditions at AdS infinity.  In all known examples where a macroscopic spacetime and gravitational physics emerge from gauge theory, the local dynamics is given by string theory.  This is consistent with the lore that string theory has no free parameters, the local dynamical laws are completely fixed.  This was the conclusion when string theory was first constructed as an expansion around a fixed spacetime, and it has not been altered as the theory has been rediscovered in various dual forms; it is one of the principal reasons for the theory's appeal.

Let us also relax the other assumptions from the introduction, large 't Hooft coupling and large $N_c$.  The AdS radius $\ell = (g_{\rm YM}^2 N_c)^{1/4} \ell_s \sim N_c^{1/4} G^{1/8}$ becomes small compared to the string size when the 't Hooft coupling is small, and comparable to the Planck scale when $N_c$ is not large.  This is consistent with our argument that we needed strong coupling and large $N_c$ in order to see macroscopic gravity.  However, string theory remains well-defined on spaces of large curvature, so the string dual should still make sense; hence our assertion that even the strong and weak nuclear interactions can be written as string theories, though in highly curved spaces.\footnote{There have been proposals that a five-dimensional picture is phenomenologically useful even for real QCD; see the recent papers Erlich \mbox{\it et~al.} (2006), Brodsky \& de Teramond (2006), and references therein.}

In more detail, consider first varying the 't Hooft coupling.  The string world-sheet action in $AdS_5 \times S^5$ is proportional to $\ell^2/\ell_s^2 = (g_{\rm YM}^2 N_c)^{1/2}$.  This is large when the 't Hooft coupling is large, so the world-sheet path integral is then nearly gaussian (i.e.~weakly coupled).  On the other hand when the 't Hooft coupling is small the string world-sheet theory is strongly coupled: the cost of living on a space of high curvature is strong world-sheet coupling.  This limits one's ability to calculate, though in the case of $AdS_5 \times S^5$ there is enough symmetry that one might ultimately be able to solve the world-sheet theory completely (Berkovits, 2005).

Now consider varying $N_c$.  From eq.~(\ref{adsplanck}) the gauge theory expansion parameter $1/N_c^2$ matches the gravitational loop expansion parameter $G$, so we can expect an order-by-order matching.  In fact, there are various indications that the duality remains true even at finite values of $N_c$, and not just as an expansion in $1/N_c^2$.  A striking example is the `string exclusion principle' (Maldacena \& Strominger, 1998). We have noted that the wavefunctions of the gravity states on $S^5$ arises in the gauge theory from traces of products of the $\varphi^i$.  However, these fields are $N_c \times N_c$ matrices, so the traces cease to be independent for products of more than $N_c$ fields: there is an upper bound 
\begin{equation}
J / N_c \leq 1
\end{equation}
for the angular momentum on $S^5$.  From the point of view of supergravity this is mysterious, because the spherical harmonics extend to arbitrary $J$.  However, there is an elegant resolution in string theory (McGreevy \mbox{\it et al.}, 2000).  A graviton moving sufficiently rapidly on $S^5$ will blow up into a spherical D3-brane (this growth with energy is a characteristic property of holographic theories), and $J = N_c$ is the largest D3-brane that will fit in the spacetime.  Thus the same bound is found on both sides of the duality, and this is a nonperturbative statement in $N_c$: it would be trivial in a power series expansion around $1/N_c = 0$.

\subsection{Open questions}

An obvious question is, to what extent is the AdS/CFT duality proven?  

We should first note that this duality is itself our most precise definition of string theory, giving an exact construction of the theory with $AdS_5 \times S^5$ boundary conditions or the various generalizations described above.
This does not mean that the duality is a tautology, because we have a great deal of independent information about string theory, such as its spectrum, its low energy gravitational action, the weak coupling expansion of its amplitudes, and so on: the gauge theory must correctly reproduce these.  Thus the duality implies a large number of precise statements, for example about the amplitudes in the strongly coupled gauge theory at each order in $1/N_c$ and $1/\sqrt{g_{\rm YM}^2 N_c}$.\footnote{We should note that there are also purely field theoretic dualities, 
where both sides presumably have a precise definition, and whose status is very similar to that of AdS/CFT duality.  The simplest example again involves
the $D=4$, ${\cal N}=4$, Yang-Mills theory but in a different part of its parameter space, $g_{\rm YM}^2 \to \infty$ at fixed $N_c$.  The Maldacena duality relates  this field-theoretic duality to the $S$-duality of the IIB string theory.}

What has been proven is much less.  The original Maldacena argument above makes the duality very plausible but of course makes many assumptions.  The quantitative tests are largely restricted to those quantities that are required by supersymmetry to be independent of the coupling.  This is not to say that the agreement follows from supersymmetry alone.  For example, supersymmetry requires the states to lie in multiplets, but the number of multiplets (as a function of their $SO(6)$ charges) is not fixed, and the fact that it agrees for each value of the charges is a strong dynamical statement --- recall in particular that the string exclusion principle must enter to make the range of charges match.

In many ways the more impressive tests are the more qualitative ones.  The point has often been made that the claim that a ten-dimensional string theory is the same a four-dimensional field theory is so audacious that if it were incorrect this should be easy to show.  Instead we find, as we look at a wide variety of situations, that the qualitative physics is exactly what we would expect.  We have noted some of these situations above: the appearance of string-like states in the gauge theory at large boost, the matching of the confining transition with the Hawking-Page transition and with the correct $N_c$ scaling on each side, the hydrodynamic properties, the matching of the deformed geometries with the RG flows and the expected low energy physics be it conformal, massive, confining, chiral symmetry-breaking, and so on.  For the confining theories, with all conformal and supersymmetries broken, one can calculate the results of high energy scattering processes.  The results differ from QCD because the theory is different, but the differences are qualitatively just those that would be expected (Polchinski \& Strassler, 2003).

Finally, we mention a very different kind of quantitative test.  Statements about strongly coupled gauge theory can be tested directly by simulation of the theory.  The range of tests is limited by the computational difficulty, but some positive results have been reported (Antonuccio \mbox{\it et~al.}, 1999, and Hiller \mbox{\it et~al.}, 2005).

In summary, we see convincing reason to place AdS/CFT duality in the category of true but not proven.  Indeed, we regard it on much the same footing as such mathematical conjectures as the Riemann hypothesis.  Both provide unexpected connections between seemingly different structures (and speaking as physicists we find a connection between gauge theory and gravity even more fascinating than one between prime numbers and analytic functions), and each has resisted either proof or disproof in spite of concentrated attention.  In either case it may be that the final proof will be narrow and uninstructive, but it seems more likely that the absence of a proof points to the existence of important new concepts to be found.

As another open question, the dictionary relating spacetime concepts in the bulk and field theory concepts on the boundary
is very incomplete, and still being developed.  For example, while we know how to translate certain states of the CFT into bulk geometries, we do not yet know the general condition on the
state in order for a semiclassical spacetime to be well defined.  

A related issue is a more precise understanding of the conservation of information in black hole decay.  The AdS/CFT duality implies that we can find an S-matrix by passing to the gauge theory variables, but there should be some prescription directly in the gravitational theory.  The black hole information problem can be understood as a conflict between quantum mechanics and locality.  In the context of emergent spacetime it is not surprising that it is locality that yields, but we would like to understand the precise manner in which it does so.

A big open question is how to extend all this from AdS boundary conditions to spacetimes that are more relevant to nature; we did find some generalizations, but they all have a causal structure similar to that of AdS.  Again, the goal is a precisely defined nonperturbative construction of the theory, presumably with the same features of emergence that we have found in the AdS/CFT case.
A natural next step might seem to be de Sitter space.  There were some attempts along these lines, for example Strominger (2001) and Witten (2001), but there are  also general arguments that this idea is problematic (Susskind, 2003).  In fact, this may be the wrong question, as constructions of de Sitter vacua in string theory (beginning with Silverstein, 2001, and Kachru \mbox{\it et~al.}, 2003) always seem to produce states that are only metastable (see Giddings, 2003, for further discussion, and Banks, 2005, for an alternate view).  As a result, cosmology will produce a chaotic state with bubbles of all possible metastable vacua (Bousso \& Polchinski, 2000).  The question is then the nonperturbative construction of states of this kind.  The only obvious spacetime boundaries are in the infinite future, in eternal bubbles of zero cosmological constant  (and possibly similar boundaries in the infinite past).  By analogy these would be the location of the holographic dual variables (Susskind, 2003).

In conclusion, the embedding of quantum gravity in ordinary gauge theory is a remarkable and unexpected property of the mathematical structures underlying theoretical physics.  We find it difficult to believe that nature does not make use of it, but the precise way in which it does so remains to be discovered.
\vskip 1.5cm
\centerline{\bf Acknowledgments}
\vskip .2cm
This work was supported in part by NSF grants PHY99-07949, PHY02-44764, and PHY04-56556.

\end{document}